\documentclass[12pt]{article}
\usepackage{a4}
\usepackage{amssymb}
\usepackage{graphicx}
\usepackage{cite}
\usepackage{setspace}
\begin{document}

\begin{titlepage}
\renewcommand{\thefootnote}{\fnsymbol{footnote}}
\setcounter{footnote}{1}

\begin{flushright}
\begin{tabular}{l}
RIKEN-TH-208\\
RIKEN-QHP-147
\end{tabular}
\end{flushright}

\bigskip

\begin{center}
\Large \bf
Sine-Square Deformation and its Relevance to String Theory
\end{center}

\bigskip

\begin{center}
{\large Tsukasa \textsc{Tada}\footnote{e-mail:
        tada@riken.jp}
}        
\end{center}

\begin{center}
\textit{RIKEN Nishina Center for Accelerator-based Science, \\
Wako, Saitama 351-0198, Japan}
\end{center}

\bigskip

\bigskip

\begin{abstract}
Sine-square deformation,  a recently found modulation of the coupling strength in certain statistical models, is discussed in  the context of two-dimensional conformal field theories, with particular attention to open/closed string duality. This deformation is shown to be non-trivial and leads to  a divergence in the worldsheet metric. The structure of the vacua of the deformed theory is also investigated.  The approach advocated here may provide an understanding of string duality through the worldsheet dynamics.
\end{abstract}
\bigskip\bigskip
\begin{center}
\textit{Version published as Mod. Phys. Lett. A, Vol. 30, No. 19 (2015) 1550092.}
\end{center}
\renewcommand{\thefootnote}{\arabic{footnote}}
\setcounter{footnote}{0}
\end{titlepage}

\section{Introduction}

In physics, the boundary condition is often treated as a secondary issue. Similar to the term itself, boundary conditions remain peripheral, never central. In string theory, however, the boundary condition plays a fundamental role. Certain boundary conditions of the worldsheet  exhibit non-perturbative aspects of string theory through D-branes. They also distinguish between open and closed strings, which  correspond to gauge theories and gravity, respectively. This short note concerns the boundary conditions in string theory.

In recent studies of a certain class of quantum systems, systems with closed and open boundary conditions were found to have identical vacua provided that the coupling constants of the open-boundary system  are modulated in a way called {\it sine-square deformation} (SSD) \cite{SSD,SSD2}. In particular, SSD works for two-dimensional conformal field theories, which describe the worldsheets of string theory \cite{Katsura:2011ss}. Therefore, the implications of this discovery to string theory are potentially vast. 

The spatial modulation of the coupling constant is seldom investigated in condensed matter physics. However, such modulation may correspond to introducing a metric with non-trivial curvature. In this sense, the above-mentioned uncovering can be interpreted as an effect caused by the worldsheet metric. Thus, by investigating the effect of the SSD on the worldsheet, we may better understand the non-perturbative aspects of string theory such as D-branes or open/closed duality through interchanges of the boundary condition caused by the worldsheet metric. Specifically, if certain worldsheet metrics can alter the boundary condition, resulting in D-brane emission or transitions between open and closed strings, then non-perturbative aspects of string theory can be understood from the dynamics of the worldsheet through its condensation. Although, the worldsheet metric can be gauged away in the perturbative treatment of string theory, the metric may couple to the dynamics when non-perturbative effects are incorporated.

The boundary condition, by nature, stipulates the development of a system, not the other way around.  Once set up, a system only evolves within its pre-determined boundary. Therefore, if non-perturbative effects of string theory  are depicted in terms of boundary conditions, they remain unaltered throughout the system development. Here we explore a possibility that the condensation of the worldsheet metric effectively alter the boundary condition, thereby exhibiting non-perturbative effects of string dynamics. If this is the case, non-perturbative aspects of string theory can be understood in terms of  worldsheet metric dynamics. In fact, one could argue that this has been somewhat achieved by the research through matrix models  \cite{Hanada:2004im},  in which the effects of D-branes were identified.  Noting that the matrix models  are nothing but the statistical mechanics of the discretized worldsheet, here we rather seek a continuum treatment of the world sheet based on the SSD.

In this note, we attempt to clarify the role of the world sheet metric in the non-perturabative dynamics of string theory. To this end, we explore  the consequences of the SSD on the worldsheet. SSD is briefly overviewed in Section \ref{sec:ssd}, and it is applied to conformal field theory in subsection \ref{sec:sl2cvac}. In subsection \ref{subsec:nont}, we verify that the SSD is actually a non-trivial transformation. A novel state in the deformed system is presented in subsection \ref{sec:anovac}. We examine the SSD in the Lagrangian formalism in Section \ref{sec:ssdstring}. Here, we reveal large divergence of the worldsheet metric. We conclude with notes and future perspectives in Section \ref{sec:sum}.

\section{Sine-square deformation}\label{sec:ssd}
First, we explain the SSD introduced by Gendiar,  Krcmar and Nishino \cite{SSD,SSD2}. Consider a system  of $N$ quantum operators $\sigma_n$. The operators are aligned  one-dimensionally and each is connected to the next neighbors with the strength $J_{n,n+1}$. The Hamiltonian of such a system is given by
\begin{equation}
{\cal H}_{0}=-\sum J_{n,n+1}\left( \boldmath\sigma_{n} \cdot \sigma_{n+1
}\right). \label{h0general}
\end{equation}
\begin{figure}[tbh]
\begin{center}\includegraphics[width=12cm]{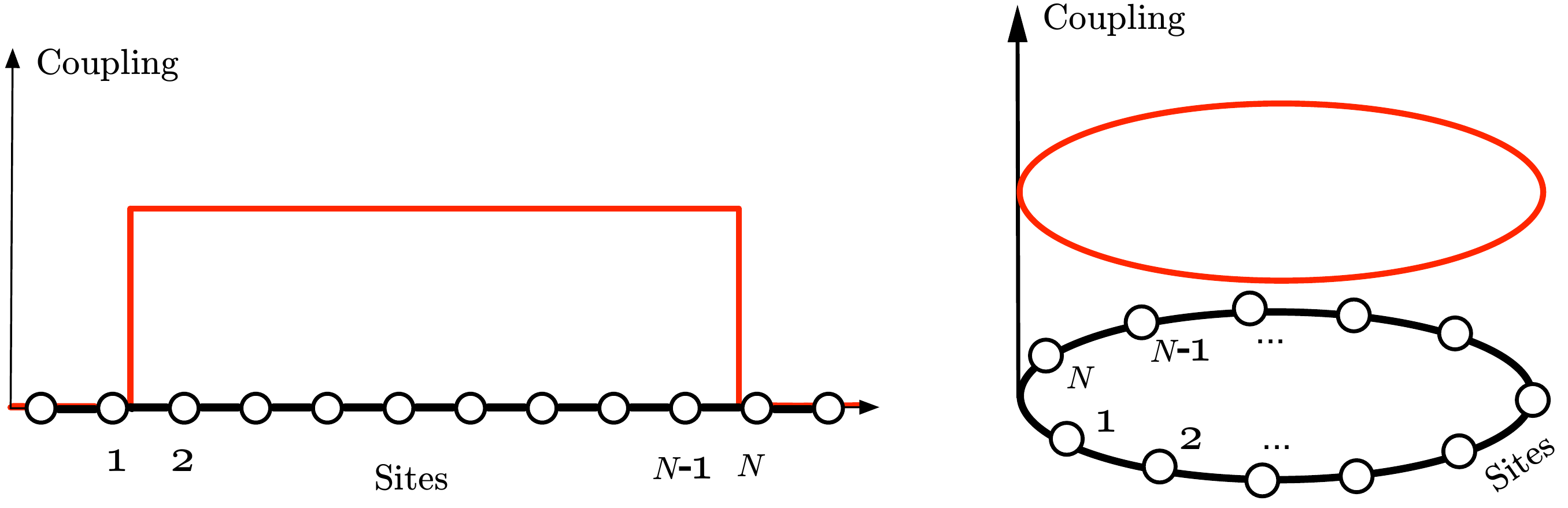}\end{center}
\caption{Open and closed boundary condition for the one-dimensional quantum system.}
\label{f2}
\end{figure}
The boundary condition of the system is governed by the configuration of the couplings $J_{n,n+1}$. Setting $J_{0,1}=J_{N,N+1}=0$ with $J_{1,2}=J_{2,3}=\cdots=J_{N-1,N}\equiv J$, the system retains an open-boundary condition. On the other hand, if $J_{N, 1}=J_{1,2}=J_{2,3}=\cdots=J_{N-1,N}\equiv J$, a closed-boundary condition is imposed (Fig. \ref{f2}). 

\begin{figure}[tbh]
\begin{center}\includegraphics[width=7cm]{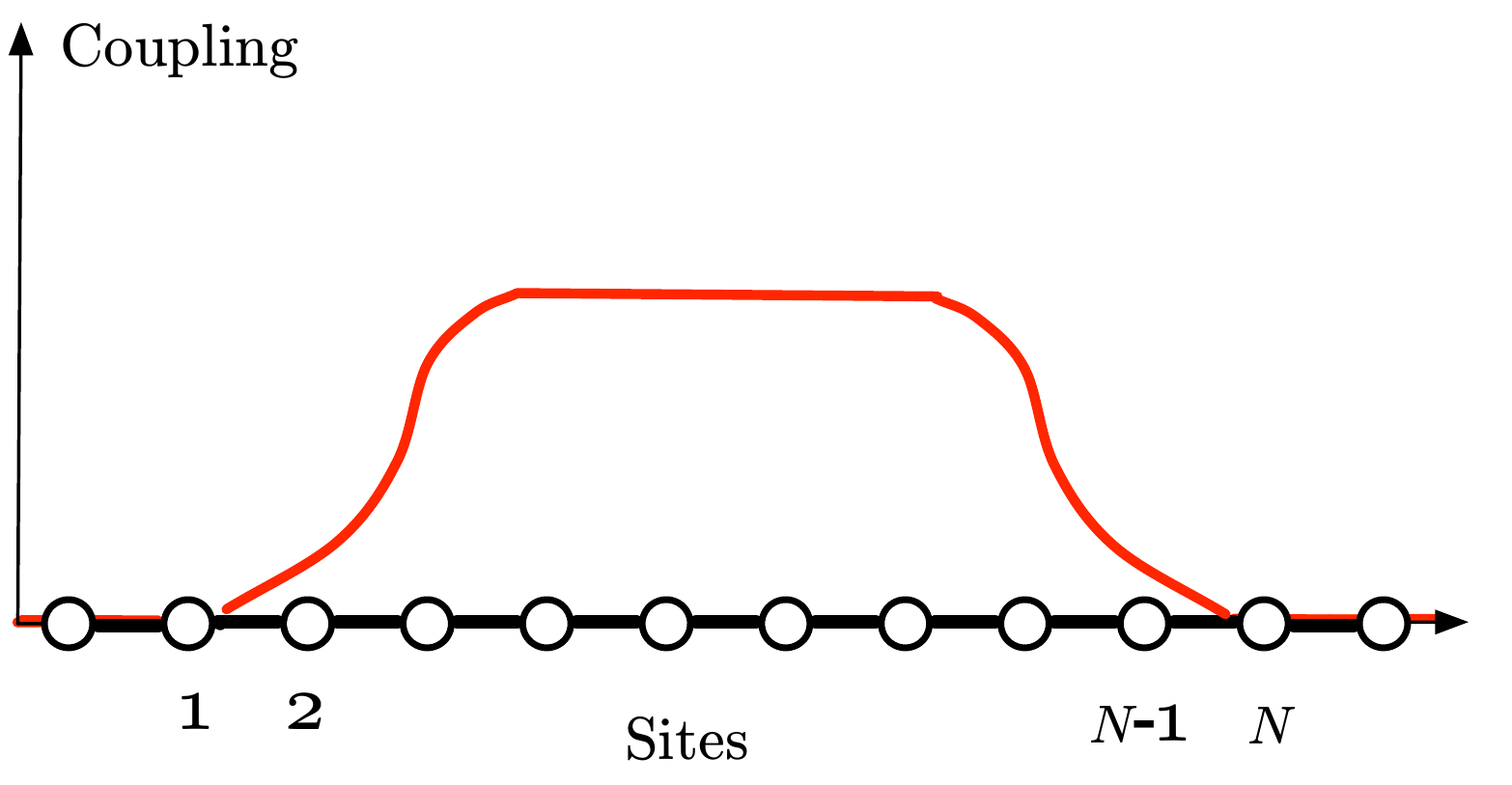}\end{center}
\caption{Modulation of the coupling is expected to reduce the edge effect.}
\label{frem}
\end{figure}

Now suppose that  the couplings are configured so that they gradually vary.  For example, in the open boundary system, we may reduce the strength of the couplings for the connections near both ends of the system (Fig. \ref{frem}).  We refer to this type of spatial coupling variation as modulation. Such modulation (illustrated in Figure \ref{frem})  is motivated by the expectation that it reduces repercussions arising from the open boundaries. However, how far we should extend the coupling modulation remains an interesting question.

\begin{figure}[tbh]
\begin{center}\includegraphics[width=14cm]{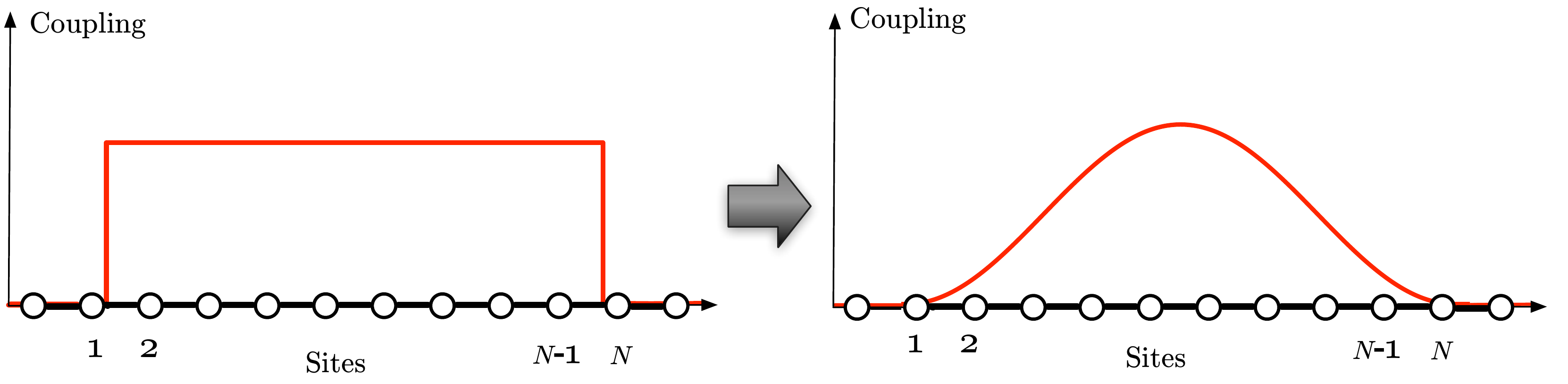}\end{center}
\caption{Sine-square deformation of the coupling for the one-dimensional quantum system.}
\label{f3}
\end{figure}

In \cite{SSD,SSD2}, this step was boldly extended  to the midpoint (see Fig. \ref{f3}).  
This modulation is expressed as
\begin{equation}
J_{n,n+1}\equiv J\sin^2\left(\frac{n}{N}\pi\right). \label{ssddef}
\end{equation}
Note that the couplings $J_{0,1}$ and $J_{N,N+1}$ located at the both ends are retained at $0$. Therefore, the system remains an 
open-boundary system, but its couplings are modulated by (\ref{ssddef}). For obvious reasons, modulation (\ref{ssddef}) is called sine-square deformation (SSD).

An astonishing feature of SSD is that it permits a ground state of the modulated  system that is identical to the ground state of a closed-boundary system. Namely, the ground state of a certain class of quantum operators coupled by (\ref{ssddef}) coincides with that of the system with $J_{N, 1}=J_{1,2}=J_{2,3}=\cdots=J_{N-1,N}\equiv J$. Given the apparently very different topologies of these configurations, this is a remarkable result. 

This exact match between the ground states of the SSD system and that of the closed-boundary system is observed in spin-$\frac12$ XY spin-chain, 1D free fermion systems, 2D conformal field theories and 2D super conformal field theories \cite{Katsura:2011ss}. This phenomenon is also expected in spin-$\frac12$ XXZ spin-chain \cite{HikiharaNishinio:2011}, extended Hubbard model\cite{Gendiaretal2011} and Kondo lattice model \cite{ShibataHotta2011}. 

The underlying mechanism of the phenomenon can be understood as follows \cite{Proof1, Proof2}.
In additions to the original Hamiltonian (\ref{h0general}) under the closed boundary condition 
\begin{equation}
{\cal H}_{0}=-\sum_{n=1}^{N} J\left( \boldmath\sigma_{n} \cdot \sigma_{n+1
}\right) , \ \sigma_{N+1} \equiv \sigma_{1}, \label{hclosed}
\end{equation}
we introduce two ``Hamiltonians'':
\begin{equation}
{\cal H}_{\pm}=-\sum_{n=1}^{N} e^{\pm  2\pi i n/N}J\left( \boldmath\sigma_{n} \cdot \sigma_{n+1
}\right). \label{hpmgeneral}
\end{equation}
The SSD can then be formulated by replacing the original $H_{0}$ in (\ref{hclosed}) with the  new Hamiltonian:
\begin{equation}
{\cal H}_{\hbox{\tiny SSD}} \equiv \frac12 {\cal H}_{0} -\frac14 \left( {\cal H}_{+}+{\cal H}_{-}\right).
\end{equation}
Indeed,
\begin{eqnarray}
\frac12 {\cal H}_{0} -\frac14 \left( {\cal H}_{+}+{\cal H}_{-}\right)&=&-\sum_{n=1}^{N}\frac12 \left(1-\frac12 e^{2\pi  i n /N}- \frac12 e^{- 2\pi i  n/N}\right)J\left( \boldmath\sigma_{n} \cdot \sigma_{n+1
}\right) \nonumber \\
&=&-\sum_{n=1}^{N}\sin^{2}\left(2\pi\frac{n}{N}\right)J\left( \boldmath\sigma_{n} \cdot \sigma_{n+1
}\right). \label{ssdspin}
\end{eqnarray}
The $\sin^{2}\left(2\pi\frac{n}{N}\right)$ factor in (\ref{ssdspin}) clearly implies an open boundary for the sine-square deformed Hamiltonian ${\cal H}_{\hbox{\tiny SSD}} $. This openness detaches the coupling between the operators at both ends, $\sigma_{1}$ and $\sigma_{N}$.

Denoting the ground state of the original Hamiltonian by $|0\rangle$, we have
\begin{equation}
{\cal H}_{0}|0\rangle = E_{0}|0\rangle,
\end{equation}
where $E_{0}$ is the ground energy. In certain systems, $ {\cal H}_{\pm}$ annihilates the ground state of the original Hamiltonian  \cite{Katsura:2011ss},
\begin{equation}
{\cal H}_{\pm}|0\rangle = 0,
\end{equation}
yielding
\begin{equation}
{\cal H}_{\hbox{\tiny SSD}} |0\rangle = \frac12 E_{0}|0\rangle.
\end{equation}
If the energy spectrum of ${\cal H}_{\hbox{\tiny SSD}}$ can be shown to be bounded below as for 1D fermions and certain conformal field theories (CFTs), the ground state $|0\rangle$ is obviously an exact ground state of  ${\cal H}_{\hbox{\tiny SSD}}$. In some cases, we can directly argue that ${\cal H}_{\hbox{\tiny SSD}}$ has a unique ground state that corresponds to $|0\rangle$ \cite{Proof1, Proof2}.

\section{CFT and sine-square deformation}
\subsection{$SL(2, \mathbb{C})$ invariant vacuum}\label{sec:sl2cvac}
In this subsection, the SSD and its mechanism are further explained in the context of 2D CFT. Following \cite{Katsura:2011ss}, we first express the Hamiltonian of a CFT on a cylinder of circumference $L$ in terms of the energy momentum tensor with the cylindrical coordinate $w=\tau+i x =\frac{L}{2\pi}\log z$:
\begin{equation}
{\cal H}_{0}=\int^{L}_0 \frac{dx}{2\pi}\left(T_{\hbox{\tiny cyl}}(w)+{\bar T}_{\hbox{\tiny cyl}}(w)\right),
\end{equation}
where
\begin{equation}
T_{\hbox{\tiny cyl}}(w)=\left(\frac{2\pi}{L}\right)^2\left[ T(z)z^2 -\frac{c}{24}\right].
\end{equation}
The energy momentum tensor $T(z)$ comprises Virasoro generators $L_n$ as $T(z)=\sum z^{-n-2}L_n$. Thus, the Hamiltonian of the CFT can also be expressed as\begin{equation}
{\cal H}_0=\frac{2\pi}{L}\left(L_0+{\bar L}_0 \right) -\frac{\pi c}{6 L}. \label{H0}
\end{equation}

As in the previous section, we introduce ${\cal H}_{\pm}$,
\begin{equation}
{\cal H}_\pm\equiv\int^L_0 \frac{dx}{2\pi}\left( e^{\pm \frac{2\pi}{L}w}T_{\hbox{\tiny cyl}}(w)+e^{\mp \frac{2\pi}{L}{\bar w}} {\bar T}_{\hbox{\tiny cyl}}({\bar w}) \right) =\frac{2\pi}{L}\left(L_{\pm 1}+{\bar L}_{\mp 1}\right).
\end{equation}
Note that, for 2d CFTs, ${\cal H}_{\pm}$ can be written as a  linear combination of familiar Virasoro operators $L_{\pm1},{ \bar L}_{\pm1}$. ${\cal H}_{\hbox{\tiny SSD}}$  now reads
\begin{equation}
{\cal H}_{\hbox{\tiny SSD}}=\frac12{\cal H}_0 -\frac14\left({\cal H}_+ +{\cal H}_-\right) = \frac{\pi}{L}\left( L_0+{\bar L}_0-\frac{L_1+L_{-1}+{\bar L}_1+{\bar L}_{-1}}{2}\right) -\frac{\pi c}{12L}. \label{Hssd}
\end{equation}

As for the ground state of 2D CFT, it is natural to assume the  $SL(2, \mathbb{C})$ invariance. Denoting the $SL(2, \mathbb{C})$ invariant vacuum $|0\rangle$, we require that  %$T(z)|0\rangle$ is regular at $z=0\  (t\rightarrow -\infty)$; equivalently, that 
$|0\rangle$ is invariant under the global conformal transformations  generated by
$L_0, L_{\pm 1}, {\bar L}_0,{\bar L}_{\pm 1}$.  From (\ref{H0}), it follows that 
\begin{equation}
{\cal H}_0|0\rangle=E_0|0\rangle,
\end{equation}
with $E_0 = -\frac{\pi c}{6L}$. From (\ref{Hssd}),  we observe that  
$|0\rangle$ is also a ground state of ${\cal H}_{\hbox{\tiny SSD}}$, with half the energy $E_{0}$
\begin{equation}
{\cal H}_{\hbox{\tiny SSD}}|0\rangle=\frac{E_0}{2}|0\rangle,
\end{equation}
because $|0\rangle$ is annihilated not only by $L_0$ and ${\bar L}_{0}$ but also by $ L_{\pm 1}$ and ${\bar L}_{\pm 1}$. 
This analysis demonstrates the SSD mechanism in the more familiar setting of 2D conformal field theories.

\subsection{Non-triviality of ${\cal H}_{\hbox{\tiny SSD}}$}\label{subsec:nont}
At this point, it would be reasonable to question the non-triviality of ${\cal H}_{\hbox{\tiny SSD}}$. This Hamiltonian differs  from the original ${\cal H}_{0}$ only by the generators of the global conformal transformations, $ L_{\pm 1},{\bar L}_{\pm 1}$. Since  the vacuum is assumed to be invariant under the $SL(2, \mathbb{C})$-transformations or global conformal transformations, a Hamiltonian or any other operator can be modified by the $SL(2, \mathbb{C})$-transformations with no physical consequences.
In fact, we apply the following $SL(2, \mathbb{C})$- transformation to (the %holonomic part of) #by H. Katsura
holomorphic part of)${\cal H}_0$ to obtain
\begin{equation}
e^{-a\frac{L_1 - L_{-1}}{2}}L_0e^{a\frac{L_1 - L_{-1}}{2}} = \cosh a L_0 - \sinh a \frac{L_1 + L_{-1}}{2}.\label{L0toLpm}
\end{equation}
The above result appears similar to the right-hand side of (\ref{Hssd}). If ${\cal H}_{\hbox{\tiny SSD}}$ can be obtained from ${\cal H}_{0}$ by the $SL(2, \mathbb{C})$-transformations, the matching of the ground states is trivial. Closer inspection reveals that this is not the case.

If the right-hand side of (\ref{L0toLpm}) accords with ${\cal H}_{\hbox{\tiny SSD}}$,  we need to require $\cosh a=\sinh a$,  which directly contradicts the identity $\cosh ^2 a- \sinh^2 a=1$. One may take the limit as $a\rightarrow \infty$ and  suitably rescale; however, in any case,  ${\cal H}_{\hbox{\tiny SSD}}$ and ${\cal H}_0$ are not connected through the ordinary $SL(2, \mathbb{C})$-transformation.

This result can also be generally confirmed by considering the two-dimensional representations of the generators:
\begin{equation}
L_0=\frac12\left(\begin{array}{cc}1 & 0 \\0 & -1\end{array}\right), \ \ L_\pm\equiv \frac{L_1\pm L_{-1}}{2}=\frac12\left(\begin{array}{cc}0 & \pm i \\i & 0\end{array}\right).
\end{equation}
A $SL(2, \mathbb{C})$ group element is non-unitarily represented  by the products of the exponents of the following generators:
\begin{equation}
\exp (i\theta_0 L_0), \ \ \exp (i\theta_+ L_+), \  \  \exp (\theta_- L_-),
\end{equation}
which multiply to yield:
\begin{equation}
\left(\begin{array}{cc}\alpha& \beta \\ \beta^*& \alpha^* \end{array}\right), \ \ \hbox{where} \ \ |\alpha|^2- |\beta|^2=1. \label{sl2rep}
\end{equation}
In this representation, the $SL(2, \mathbb{C})$ group acts on $L_0$ as follows
\begin{equation}
\left(\begin{array}{cc}\alpha & \beta \\ \beta^* & \alpha^* \end{array}\right)^{-1} L_0 \left(\begin{array}{cc}\alpha & \beta \\ \beta^* & \alpha^* \end{array}\right) =\frac12\left(\begin{array}{cc}|\alpha|^2+|\beta|^2 & 2\alpha^* \beta \\-2\alpha \beta ^*& -|\alpha|^2-|\beta|^2\end{array}\right). \label{L0sl2ed}
\end{equation}
We now  explore the parameter region in which the above expression can be expressed as a linear combination of $L_{0}$ and $L_{+}$;
\begin{equation}
aL_{0}%+ modification
-bL_{+}=\frac12 \left(\begin{array}{cc}a & -%
ib \\-%
ib & -a\end{array}\right).\label{L0L+lincomb}
\end{equation}
Combining (\ref{sl2rep}), (\ref {L0sl2ed}) and (\ref{L0L+lincomb}), the following conditions are easily obtained:
\begin{equation}
\left\{\begin{array}{l} |\alpha|^2+|\beta|^2 =a\\2\alpha^* \beta=-2\alpha \beta^*=-%
ib \\|\alpha|^2- |\beta|^2=1\end{array}\right. \ \ . \label{abcond}
\end{equation}
The left-hand side of the following identity inequality
\begin{equation}
|\alpha %+ corrected
-i\beta|^{2} \geq 0,
\end{equation}
can be expanded as
\begin{equation}
|\alpha %+
-i\beta|^{2} =(\alpha %+
-i\beta)(\alpha^{*} %-
+i\beta^{*} )=|\alpha|^{2}+|\beta|^{2}%+
-i\alpha^{*}\beta%-
+i\alpha \beta^{*},
\end{equation}
yielding
\begin{equation}
a-b \geq 0, \label{abqeq0}
\end{equation}
where we have used the first and second conditions in (\ref{abcond}). The case of interest is $a=b$,  which implies that a $SL(2, \mathbb{C})$ action on $L_0$ or ${\cal H}_0$ yields ${\cal H}_{\tiny SSD}$ up to a normalization. However,  (\ref{abqeq0}) becomes an equality only when $\alpha=%-
i\beta$, which directly contradicts the third condition in (\ref{abcond}). Therefore, we have proven by contradiction that  $SL(2, \mathbb{C})$ cannot act on ${\cal H}_0$  to yield ${\cal H}_{\tiny SSD}$.
\footnote{
%Revision 09/09/2014
% transformation from $L_0$ to $L_0-(L_{1}+L_{-1})/2$ can only be evoked by the highly singular coordinate %transformation of the Riemann surface t $w=e^{\frac{2}{z-1}}$. \cite{IshibashiTada}\ However,
%
Alexandros Kehagias has alerted the author that $U_{q}(sl(2))$ action on $H_{0}$ might yield ${\cal H}_{\tiny SSD}$.} \label{footnote:uqsl2}

%Revision 09/16/2014
In fact, we can construct the transformation from $L_0$ to $L_0-(L_{1}+L_{-1})/2$ on the Riemann surface explicitly and confirm that it lies outside $SL(2, \mathbb{C})$, as follows.
Since Virasoro operators can be expressed on the Riemann surface as $L_{n}=z^{n+1}\partial_{z}$,
\begin{equation}
L_0-(L_{1}+L_{-1})/2 = \frac12(2z-1-z^{2})\partial_{z} = -\frac12(z-1)^{2}\partial_{z}, \label{eq:ssdz}
\end{equation}
while 
\begin{equation}
L_0 = \ u\partial_{u}, \label{eq:l0u}
\end{equation}
in a different complex variable on the Riemann surface. Then, the transformation takes $L_0$ to $L_0-(L_{1}+L_{-1})/2$ is nothing but the transformation from $z$ to $u$, $u=f(z)$ that equates Eq. (\ref{eq:ssdz}) with Eq. (\ref{eq:l0u}). One finds the explicit form of the transformation as
\begin{equation}
u=e^{\frac{2}{z-1}}, \label{eq:uz}
\end{equation}
which contains an essential singularity.  The transformation (\ref{eq:uz}) is obviously different from the 
$SL(2, \mathbb{C})$ transformation, which should have been expressed as $u=\frac{az+b}{cz+d}$. Further analysis utilizing the explicit formula (\ref{eq:uz}) will be reported in future publication.

\subsection{Another vacuum?}\label{sec:anovac}
In subsection \ref{sec:sl2cvac}, the $SL(2, \mathbb{C})$ invariant vacuum $|0\rangle$ was shown to also constitute  the lowest-energy $\frac12 E_{0}$ eigenstate of ${\cal H}_{\hbox{\tiny SSD}}$. Thus, we may naturally seek other eigenstate of  ${\cal H}_{\hbox{\tiny SSD}}$. 
For the original Hamiltonian ${\cal H}_{0}$, there exists a set of eigenstates corresponding to primary fields of CFT:
\begin{equation}
|h, {\bar h}\rangle \equiv \phi(0,0) |0\rangle,
\end{equation}
where $\phi$ is a primary field of dimensions $h$ and $\bar h$. Unlike $|0\rangle$, $|h, {\bar h}\rangle$ is not an eigenstate of ${\cal H}_{\hbox{\tiny SSD}}$, but provides a useful starting point. Consider a state of the form
\begin{equation}
\sum_{n} a_{n} \left( L_{{-1}} \right)^{n}|h\rangle, \label{eqn:seriesh}
\end{equation}
where we have focused on the holomorphic part for simplicity.
Simple calculation shows that for (\ref{eqn:seriesh}) to be an eigenstate of ${\cal H}_{\hbox{\tiny SSD}}$,
\begin{equation}
\left(L_{0}-\frac{L_{1}+L_{-1}}{2}\right)\sum_{n} a_{n} \left( L_{{-1}} \right)^{n}|h\rangle=\alpha \sum_{n} a_{n} \left( L_{{-1}} \right)^{n}|h\rangle,
\end{equation}
the following recurrence relation for $a_{n}$ should hold:
\begin{equation}
\left( 2\left(n+1\right) h +n \left( n+1\right)\right) a_{n+1} +2 \left( \alpha -n-h\right) a_{n} +a_{n-1} =0.
\end{equation}
A solution to the above recurrence relation is 
\begin{equation}
a_{n}=\frac{1}{n!}, \ \  \alpha=0.
\end{equation}
Thus, we appear to have identified vacuums other than $|0\rangle$ of the form
\begin{equation}
e^{L_{-1}}|h\rangle. \label{evv}
\end{equation}
However, since the norm of (\ref{evv}) is divergent,  a limiting process is required to properly define (\ref{evv}).
\footnote{After the completion of the manuscript, we learned from H. Katsura that  there is another non-normalizable vacuum, which takes the form of $\sum_{n>1} L_{-n}|0\rangle$.}

\section{SSD and strings}\label{sec:ssdstring}
%Added 2/27/2015
In this section, we examine the behavior of  the two-dimensional massless scalar field under SSD, as a step towards the application of SSD to string theory. For this purpose, we need to espouse the Lagrangian formalism.
In the following exposition, we adopt the notation of \cite{DiFrancesco:1997nk}.
The Lagrangian of the two-dimensional free bosonic field $\phi (x,t) $ is 
\begin{equation}
{\cal L}_0=\frac12 g \int dx\left\{ \left(\partial_t \phi \right)^2 -\left(\partial_x \phi \right)^2 \right\},
\end{equation}
where the non-dimensional normalization of the Lagrangian $g$ is left undetermined, for convenient comparison with different conventions. A common convention is $g=\frac{1}{4\pi}$. We consider the bosonic field $\phi(x,t)$ on a cylinder of circumference $L$ so that $\phi(x+L,t)=\phi(x,t)$. Then, the field is expressed by the  following Fourier expansion:
\begin{eqnarray}
\phi(x,t)&=&\sum_{n\in\mathbb{Z}}e^{2\pi i n x/L} \phi_n, \label{eq:phifourer}\\
\hbox{where} \quad \phi_n(t)&\equiv&\frac{1}{L}\int_0^L dx e^{-2\pi i n x/L}\phi(x,t).
\end{eqnarray}
In terms of the Fourier components $\phi_n(t)$, the Lagrangian is expressed as
\begin{equation}
{\cal L}_0=\frac12 g \sum_n \left\{ \dot{\phi}_n \dot{\phi}_{-n}-\left(\frac{2\pi n}{L}\right)^2 \phi_n \phi_{-n} \right\}.
\end{equation}
Then the momentum conjugate to $\phi_n$ is
\begin{equation}
\pi_n=gL\dot{\phi}_{-n}.
\end{equation}
The Hamiltonian is obtained as
\begin{equation}
{\cal H}_0=\frac{1}{2gL}\sum_n\left\{ \pi_n \pi_{-n}+\left(2\pi n g \right)^2 \phi_n \phi_{-n} \right\}. \label{eq:H0p}
\end{equation}
Note that $\phi_n^\dagger = \phi_{-n}$ and $\pi_n^\dagger = \pi_{-n}$.

Introducing 
\begin{eqnarray}
a_n &\equiv& -i n\sqrt{\pi g}  \phi_n+\frac{1}{\sqrt{4\pi g}}\pi_{-n}\ , \nonumber \\ &&\\
\bar{a}_n &\equiv& - i n \sqrt{\pi g}\phi_{-n}+\frac{1}{\sqrt{4\pi g}}\pi_{n}, \nonumber
\end{eqnarray}
we obtain the following commutation relations from the canonical commutation relation $[\phi_n, \pi_m]=i\delta_{nm}$
:
\begin{equation}
[a_n, a_m]=n\delta_{n+m}\ , \qquad [\bar{a}_n, \bar{a}_m]=n\delta_{n+m}\ , \qquad [a_n, \bar{a}_m]=0.
\end{equation}
In terms of $a_n, \bar{a}_n$, the Hamiltonian is expressed  as
\begin{equation}
{\cal H}_0=\frac{2\pi}{L}\left\{ \sum_{n>0} a_{-n}a_n+\frac12 a_0^2+\sum_{n>0}\bar{a}_{-n}\bar{a}_n +\frac12\bar{a}_0^2 \right\}.
\end{equation}
Note that $a_0=\bar{a}_0=\frac{\pi_0}{\sqrt{4\pi g}}.$

When $a_n$'s and $\bar{a}_n$'s  are treated as operators, they can form the following Virasoro operators:
\begin{eqnarray}
L_n&=&\frac12\sum_{m\in \mathbb{Z}} a_{n-m}a_m \qquad (n \neq 0)\ , \quad L_0=\sum_{n>0} a_{-n}a_n+\frac12a_0^2 \ ,\label{eq:Ln}\\
\bar{L}_n&=&\frac12\sum_{m\in \mathbb{Z}} \bar{a}_{n-m}\bar{a}_m \qquad (n \neq 0)\ , \quad \bar{L}_0=\sum_{n>0} \bar{a}_{-n}\bar{a}_n+\frac12\bar{a}_0^2 \ .\label{eq:Lbarn}
\end{eqnarray}
The Hamiltonian and  Virasoro operators are related as follows:
\begin{equation}
{\cal H}_0=\frac{2\pi}{L}\left( L_0 + \bar{L}_0 \right), \label{eq:H0L0}
\end{equation}
up to a constant, which is irrelevant in the following discussion.

We can consider new terms if the Hamiltonian contains $L_1$ and $L_{-1}$.  Equations (\ref{eq:Ln}) and (\ref{eq:Lbarn}) can be expressed in the form
\begin{equation}
L_{1}=\frac12\sum_{m\in\mathbb{Z}}a_{1-m}a_{m}=\frac12\sum_{m\in\mathbb{Z}}\left(-\sqrt{\pi g}i\left(1-m\right)\phi_{1-m}+\frac{\pi_{m-1}}{\sqrt{4\pi g}}
\right)\left(-\sqrt{\pi g}im\phi_{m}+\frac{\pi_{-m}}{\sqrt{4\pi g}}
\right),\end{equation}
from which the following relation is easily observed:
\begin{eqnarray}
\frac{2\pi}{L}\Big( L_{1}+\bar{L}_{1}&+&L_{-1}+\bar{L}_{-1}
\Big) 
=\frac{1}{2gL}\sum_{n\in\mathbb{Z}}\Big\{ \pi_{n}\pi_{-(n+1)}+\pi_{n}\pi_{-(n-1)} \nonumber \\&+&(2\pi g)^{2}n\left(n+1\right)\phi_{n}\phi_{-(n+1)}+(2\pi g)^{2}n\left(n-1\right)\phi_{n}\phi_{-(n-1)}\Big\}. \label{eq:L1Lm1}
\end{eqnarray}
We now proceed to  evaluate the  the Lagrangian expression corresponding to the deformed Hamiltonian ${\cal H}_{\hbox{\tiny SSD}} \sim L_{0}+\bar{L}_{0}-\frac12 (L_{1}+\bar{L}_{1}+L_{-1}+\bar{L}_{-1})$.

We may reasonably expect a general form of the corresponding deformed Lagrangian, such as
\begin{equation}
{\cal L}=\frac12 \int_{0}^{L} dx\left\{\left(\partial_t\varphi\right)F(x) \left(\partial_t\varphi\right)-\left(\partial_x\varphi\right)G(x)\left(\partial_x\varphi\right) \right\}. \label{eq:FGL}
\end{equation}
Postulating the following forms for $F(x)$ and $G(x) $
\begin{equation}
F(x)=N\sum_{k\in \mathbb{Z}} r^{|k|}e^{2\pi i k x/L} \ \ \hbox{and}\ \   G(x) = 1-\alpha\cos\frac{2\pi x}{L}, \label{eq:FGpos}
\end{equation}
we determine whether  the deformed Lagrangian generates $H_{\tiny SSD}$.  In (\ref{eq:FGpos}), the parameter $\alpha$  represents  deformation.  The number $r$ should be less than unity and may depend on the value of $\alpha$. $N$ is the normalization factor for $G$ and may also depend on $\alpha$. Since (\ref{eq:FGpos}) should revert to the original Lagrangian when $\alpha = 0$, we expect that $r\rightarrow 0$ and $N\rightarrow 1$ as $\alpha \rightarrow 0$. The deformed Lagrangian, denoted $L_{\alpha}$ as a reminder of the role of  $\alpha$, can be expressed in terms of the Fourier modes $\dot{\phi}$
\begin{eqnarray}
{\cal L}_{\alpha}&=&\frac12 \int_{0}^{L} dx\left\{\left(\partial_t\varphi\right)F(x) \left(\partial_t\varphi\right)-\left(\partial_x\varphi\right)G(x)\left(\partial_x\varphi\right) \right\} \nonumber \\
&=& \frac{gL}{2}\sum_{n, k}\dot{\phi}_{n}\dot{\phi}_{-n-k}Nr^{|k|} \\&&-\frac{2\pi^{2}g}{L}\left\{n^{2}\phi_{n}\phi_{-n}-\frac{\alpha}{2}\left(n\left(n+1\right)\phi_{n}\phi_{-n-1}
+n\left(n-1\right)\phi_{n}\phi_{-n+1}\right)\right\}. \nonumber
\end{eqnarray}
Introducing the following notation 
\begin{equation}
F_{nm}\equiv \sum_{k\in\mathbb{Z}}\delta_{n+m+k, 0}Nr^{|k|}, \label{eq:Fdef}
\end{equation}
the kinetic part of the Lagrangian can be simply expressed as
\begin{equation}
\frac{gL}{2}F_{nm}\dot{\phi}_{n}\dot{\phi}_{m}. \label{eqn:kinterm}
\end{equation}
From (\ref{eqn:kinterm}), it follows that the canonical conjugate momentum becomes
\begin{equation}
\pi_{n}=gL \sum_{m}  F_{nm}\dot{\phi}_{n}=gL\sum_{k} Nr^{|k|}\dot{\phi}_{-n-k}.
\end{equation}
The last equality follows from the explicit definition of $F_{mn}$ (\ref{eq:Fdef}).

Here we claim that
\begin{equation}
{}^\exists N, {}^\exists r \qquad F_{nm}^{-1} =\delta_{-n,m}-\frac{\alpha}{2}\delta_{-n-1,m}-\frac{\alpha}{2}\delta_{-n+1,m}. \label{eq:claimFminus}
\end{equation}
In other words, $N$ and $r$ can be expressed in terms of $\alpha$ such that
\begin{equation}
\frac{1}{gL}\sum_{m}\left( \delta_{-n,m}-\frac{\alpha}{2}\delta_{-n-1,m}-\frac{\alpha}{2}\delta_{-n+1,m}\right) \pi_{m}=\dot{\phi}_{n}.\label{eq:claimpiphi}
\end{equation}
To validiate claims (\ref{eq:claimFminus}) or (\ref{eq:claimpiphi}) , we require that
\begin{eqnarray*}
&&\frac{1}{gL}\sum_{m}\left( \delta_{-n,m}-\frac{\alpha}{2}\delta_{-n-1,m}-\frac{\alpha}{2}\delta_{-n+1,m}\right) \pi_{m}\\
&&=\frac{1}{gL}\sum_{m,k}\left( \delta_{-n,m}-\frac{\alpha}{2}\delta_{-n-1,m}-\frac{\alpha}{2}\delta_{-n+1,m}\right) gL Nr^{|k|}\dot{\phi}_{-m-k} \\
&&=N\sum_{k\in\mathbb{Z}}\left\{ r^{|k|} \dot{\phi}_{n-k}-\frac{\alpha}{2}r^{|k|}\dot{\phi}_{n+1-k}-\frac{\alpha}{2}r^{|k|}\dot{\phi}_{n-1-k}\right\}\\
&&=N\sum_{k\in\mathbb{Z}}\left\{ r^{|k|} \dot{\phi}_{n-k}-\frac{\alpha}{2}r^{|k-1|}\dot{\phi}_{n-k}-\frac{\alpha}{2}r^{|k+1|}\dot{\phi}_{n-k}\right\}
\end{eqnarray*}
is identical to $\dot{\phi}_{n}$. This requirement can be met only under the following conditions:
\begin{equation}
\cases{
r^{k}-\frac{\alpha}{2}r^{k-1}-\frac{\alpha}{2}r^{k+1}=0 \quad (k\geq 1) \cr
N(1-\frac{\alpha}{2}r-\frac{\alpha}{2}r)=1 \quad  (k=1) \cr
r^{-k}-\frac{\alpha}{2}r^{-k+1}-\frac{\alpha}{2}r^{-k-1}=0 \quad (k\leq -1) \cr
}
. \label{eq:Nrcond}
\end{equation}
It is trivial to see that conditions (\ref{eq:Nrcond}) are satisfied if
\begin{equation}
r-\frac{\alpha}{2}-\frac{\alpha}{2}r^{2}=0 \ , \qquad N=\frac{1}{1-\alpha r}.
\end{equation}
Solving the above quadratic equation and demanding that $r\rightarrow 0$ as $\alpha \rightarrow 0$, we find that the expressions
\begin{equation}
r=\frac{1-\sqrt{1-\alpha^{2}}}{\alpha} \ , \qquad N=\frac{1}{\sqrt{1-\alpha^{2}}} \label{eq:rNsol}
\end{equation}
validate  claims (\ref{eq:claimFminus}) and (\ref{eq:claimpiphi}). We assume that $r$ and $N$ satisfy (\ref{eq:rNsol}) and that $F_{mn}$ is accordingly determined from (\ref{eq:Fdef}) in the following.

The Hamiltonian corresponding to $L_{\alpha}$, which we denote $\cal{H}_{\alpha}$, is now calculated as
\begin{eqnarray*}
{\cal H}_{\alpha}&=&\sum_{n}\pi_{n} \dot{\phi}_{n}-{\cal L}_{\alpha} \\
&=&\sum_{n,m}\pi_{n} \frac{1}{gL}F_{mn}^{-1}\pi_{n}-\sum_{n,m}\frac12
F_{nm}\dot{\phi}_{n}\dot{\phi}_{m}\\ &&+\sum_{n}\frac{2\pi^{2}g}{L}\left\{n^{2}\phi_{n}\phi_{-n}-\frac{\alpha}{2}n\left(n+1\right)\phi_{n}\phi_{-n-1}-\frac{\alpha}{2}n\left(n-1\right)\phi_{n}\phi_{-n+1}\right\} \\
&=&\frac{1}{2gL}\sum_{n,n}\pi_{n}F_{nm}^{-1}\pi_{m}\\ &&+\sum_{n}\frac{2\pi^{2}g}{L}\left\{n^{2}\phi_{n}\phi_{-n}-\frac{\alpha}{2}n\left(n+1\right)\phi_{n}\phi_{-n-1}-\frac{\alpha}{2}n\left(n-1\right)\phi_{n}\phi_{-n+1}\right\} \\
&=&\frac{1}{2gL}\Bigl[
\pi_{n}\pi_{-n}-\frac{\alpha}{2}\pi_{n}\pi_{-n+1}-\frac{\alpha}{2}\pi_{n}\pi_{-n-1} \\
&&+\left(2\pi g \right)^{2}n^{2}
\phi_{n}\phi_{-n}-\frac{\alpha}{2}\left(2\pi g \right)^{2}n\left(n+1\right)\phi_{n}\phi_{-n-1}-\frac{\alpha}{2}\left(2\pi g \right)^{2}n\left(n-1\right)\phi_{n}\phi_{-n+1}\Bigr],
\end{eqnarray*}
which evaluates to
\begin{equation}
\frac{2\pi}{L}\left(L_{0}+ \bar{L}_{0} - \frac{\alpha}{2}\left(L_{1}+\bar{L}_{1}+L_{-1}+\bar{L}_{-1} \right) \right), \label{eq:inbet}
\end{equation}
using (\ref{eq:L1Lm1}), (\ref{eq:H0p}), and (\ref{eq:H0L0})
\footnote{A similar in-between Hamiltonian with (\ref{eq:inbet}) was also discussed in \cite{Proof2}.}.
Thus, ${\cal H}_{\alpha}$ varies from the original free Hamiltonian to a sine-square deformed Hamiltonian up to the overall  factor $\frac12$ as  $\alpha$ is varied from $0$ to $1$. When  $\alpha=1$ in (\ref{eq:rNsol}) , $r=1$ and (\ref{eq:FGpos}) becomes
\begin{equation}
F(x)=N\sum_{k\in \mathbb{Z}} e^{2\pi i k x/L} = N \delta (x) \ \ \hbox{and}\ \   G(x) = 1-\cos\frac{2\pi x}{L} .\label{eq:FGposa1}
\end{equation}
However, from (\ref{eq:rNsol}), we note that $N$ in (\ref{eq:FGposa1}) diverges as  $\alpha$ tends to unity. 

%Revision 2/20/2015
Thus,  we find that $F(x)$, the part of the  Lagrangian which is supposedly correspond to $g_{00}$ component of the worldsheet metric,  severely diverges under the SSD, at least in the gauge applied here. While one may expect a divergence upon such a singular event like the change of the boundary condition, here we have not exhausted all the options to remedy the divergence. Nonetheless, the appearance of such divergence impedes attempts at further analysis. A possible approach may be to maintain $\alpha$ away from unity. In this approach, $\alpha$ may serve as a regularization parameter. These analyses%
%Revise 09/09/2014
, as well as the quantization of the total system and the associated question of the gauge fixing,
 are left for future study.

\section{Discussion}\label{sec:sum}
To better understand the relationships between open and closed strings, we investigated the SSD of string theory. We encountered strong divergence in the worldsheet metric of the sine-square deformed model. 
This divergence in the Lagrangian could be partly caused by the continuous treatment of the worldsheet. Therefore one may try to discretize the worldsheet itself \cite{thooft}besides the approach proposed in the previous section. One such attempt could be achieved by the use of matrix models.  Another might be
an introduction of non-commutativity on the worldsheet. Non-commutative worldsheet had been considered in \cite{Chaichian:2000ba}  and it had lead the deformed Virasoro algebra \cite{Shiraishi:1995rp,Shiraishi2,Shiraishi3}. The point raised in the footnote in subsection \ref{footnote:uqsl2} might also be relevant to this respect.

%Revision 11/19/2014
Of course, the ultimate question should be, what is the nature of SSD theory. Though we started from conformal field theory,  after applying the SSD, there is no guarantee that conformal symmetry is still preserved even partially. While the retention of conformal symmetry (at least partially) is certainly desirable for SSD theory to be useful in the study of string theory, the analysis presented here is  inconclusive on the matter. One suggestive finding presented here, though, is the degenerate vacua. If there are more degenerate states we have not yet found, it is imaginable that there are also many other states whose energy eigenvalues are close to the vacuum energy. Then, it might  imply that the system possesses a continuous spectrum. This point should be pursued further in future studies.

We also emphasize that the coupling constant of our analyzed system was spatially modulated. In statistical models, such modulations have not played a significant role. However, this is exactly what we would do when one introduces gravity to the model. In the context of string theory, this rather seems to be a natural option. In fact, \cite{Kostov:2012wv} considered inhomogeneous XXX model, the particular case of XXZ model. The inhomogeneity there differs from our analysis  but it may have relevance in a wider sense. It is possible that a variety of spatial modulations introduced to statistical models (especially solvable ones) may reveal a rich structure and become essential in  future studies of string theory.

%%Addendum 03/16/2014
%{\it Note added in proof.} After the completion and the acceptance of the manuscript, we learned that the discussion in Subsec. \ref{subsec:nont} can be more sophisticated if one uses the formulae presented in Ref.~\refcite{Matone:2015wxa}.
%%

\section*{Acknowledgements} 
This study is supported  in part by JSPS KAKENHI Grant No. 25610066 and the RIKEN iTHES Project. We would like to thank N. Ishibashi for the collaboration during the early stage of the study. We also thank H. Katsura for his comments on the manuscript and valuable inputs. Useful discussions and comments by H. Itoyama, V. Kazakov, I. Kostov, Y. Matsuo and other participants of ``Todai/Riken joint workshop on Super Yang-Mills, solvable systems and related subjects'' are gratefully acknowledged. The author's  gratitude also extends to C. Ahn, A. Kehagias, K. Lee and other participants and the organizers of ``3rd Bangkok workshop on high energy theory.'' Last but not least, the author is indebted to H. Kawai and T. Yoneya for helpful comments, guidances and encouragements.

\end{document}